\documentclass[12pt]{article}
\input epsf.tex
\def\DESepsf(#1 width #2){\epsfxsize=#2 \epsfbox{#1}}
\usepackage{epsfig}
\usepackage{color}
\usepackage{array}
\makeatletter
\def\alt{\mathrel{\mathpalette\gl@align<}}
\def\agt{\mathrel{\mathpalette\gl@align>}}
\def\gl@align#1#2{\lower.6ex\vbox{\baselineskip\z@skip\lineskip\z@
\ialign{$\m@th#1\hfil##\hfil$\crcr#2\crcr\sim\crcr}}}
\makeatother

\begin{document}
\begin{flushright}
{\tt hep-ph/0507319}\\
July, 2005 \\
UMD-PP-05-047\\
\end{flushright}
\vspace*{2cm}
\begin{center}
{\baselineskip 25pt \Large{\bf
Neutrino Mixing Predictions of a Minimal SO(10) Model \\
with Suppressed Proton Decay}
}

\vspace{1cm}

{\large
Bhaskar Dutta{$^*$},
Yukihiro Mimura{$^*$}
and R.N. Mohapatra{$^\dagger$}}

\vspace{.5cm}

{\it
$^*$Department of Physics, University of Regina, \\
Regina, Saskatchewan S4S 0A2, Canada \\
$^\dagger$Department of Physics and Center for String and Particle Theory, \\
University of Maryland, College Park, MD 20742, USA\\
}
\vspace{.5cm}

\vspace{1.5cm}
{\bf Abstract}
\end{center}

During the past year, a minimal renormalizable supersymmetric
SO(10) model has been proposed with the following properties: it
predicts a naturally stable dark matter and neutrino mixing
angles $\theta_{\rm atm}$ and $\theta_{13}$ while at the same time
accommodating CKM CP violation among quarks with no SUSY CP
problem. Suppression of proton decay for all allowed values of
$\tan\beta$ strongly restricts the flavor structure of the model
making it predictive for other processes as well. We discuss
the following predictions of the model in this paper,
e.g. down-type quark masses,
and neutrino oscillation parameters,
$U_{e3}$, $\delta_{\rm MNSP}$, 
which will be tested by 
long baseline experiments such as T2K 
and subsequent experiments
using the  neutrino beam from JPARC.
We also calculate lepton flavor violation and the 
lepton asymmetry of the Universe in this model.

\thispagestyle{empty}
\bigskip
\newpage

\addtocounter{page}{-1}

\section{Introduction}
\baselineskip 20pt

The existence of  non-zero neutrino masses appears to have
considerably narrowed the choice of grand unified theories, with
those based on the group SO(10) being preferred for several
reasons. SO(10) is the minimal group with all the ingredients for
a small neutrino mass since its {\bf 16}-dimensional spinor
representation contains the right-handed neutrino, $\nu_R$, which
is needed to implement the seesaw mechanism\cite{seesaw} along
with the other fermions of the standard model. It also
 contains the $B\!-\!L$ symmetry needed
to keep the right-handed neutrino masses below the Planck scale
and provides a group theoretic rational for the belief that
neutrinos most likely are  Majorana particles. The unification of
all the fermions into the {\bf 16}-plet also raises the hope that
the number of parameters required to describe fermion masses and
mixing will be less than those in the standard model, thus making
the model predictive for neutrinos.

Initial attempts to realize this hope were made in
 a class of minimal SO(10) models with a single
{\bf 10} and a single $\overline{\bf 126}$ Higgs
multiplets\cite{Babu:1992ia,Goh:2003sy} which led to predictions for
neutrino mixings as well as two of the masses without any extra
symmetry assumptions. The gratifying result was the natural
manner in which the large solar and atmospheric mixings arose in
these models with predictions in gross agreement with current
neutrino observations. The detailed predictions for the case where
the Yukawa couplings are CP conserving are however away from the
current central values of the neutrino parameters though still in
agreement with observations at three $\sigma$ level.

Encouraged by this initial success in understanding large
neutrino mixings, attempts were made to study CP violation in
this model\cite{moha,Dutta:2004wv} by making the Yukawa couplings
complex. It was found that fitting fermion masses and mixings
forces the CKM phase to be in the second quadrant
rather than in the first. 
This question has been reanalyzed in two recent papers:
(1) using type II seesaw in these context of SO(10) models\cite{type2},
if one allowed a very high value of the strange quark mass, the CKM phase 
could be in the
first quadrant\cite{Bertolini:2005qb};
(2) in type I seesaw (and mixed case),
it is shown that
all the masses and mixings (within 99\% CL) can be fit
for a finely tuned  range of parameters\cite{babu}.
While one could
take this to be an indication possibly of new physics
contributions to CP violation, a more conservative point of view
would be to demand that the model be extended to generate CKM CP
violation and see whether its predictivity in the neutrino sector
is still preserved.  One way to achieve this is to go beyond this
minimal Higgs structure by including the {\bf 120} Higgs 
field\cite{Dutta:2004hp,Dutta:2004zh,Bertolini:2004eq}.
In particular,
it was shown in Ref.\cite{Dutta:2004hp,Dutta:2004zh} that the model
not only accommodates CKM CP violation but it also solves the SUSY
CP problem\cite{Dutta:2004hp} and has the potential to solve the
strong CP problem as well. This would make such models plausible
candidates for a theory not only of neutrino masses but also of
CP violating phenomena.

As is well known, GUT models generically lead to an unstable
proton and therefore the lifetime of proton provides additional
constraints on them. Several GUT models have already been
severely constrained by the present experimental bounds on the
proton lifetime\cite{Hisano:2000dg,Goto:1998qg}. The
question of proton decay in this class of SO(10) models has been
under scrutiny in several papers\cite{Dutta:2004zh,Goh:2003nv,Fukuyama:2004pb}. 
In particular, 
in Ref.\cite{Dutta:2004zh}, we showed that requiring the suppression of
proton decay for both small and large $\tan\beta$ severely
constrains the textures of the Yukawa couplings. In particular,
${\bf 120}$ becomes a necessity for this to happen. The reason
for this is that cancellation of different terms in the decay
amplitudes for different proton decay processes are absent in
this formalism. This makes the model very predictive. In this
paper, we study these predictions.

 In the
process of suppressing the decay amplitude, the number of free
parameters becomes less and the model becomes very predictive in
the quark-lepton sector. For example, the down-type quark masses,
$|U_{e3}|$, $\delta_{\rm MNSP}$ etc, are predictions of this model. We
also predict the lepton asymmetry which then get converted to
baryon asymmetry via sphaleron process in the thermal leptogenesis scenario.
Since the baryon
asymmetry is determined quite accurately from the recent
experimental data, it is important to calculate the prediction of
this model. We can also predict the probability of muon type
neutrino to electron type neutrino oscillation
($P_{\nu_{\mu}\rightarrow \nu_e}$) as a function of distance and
the energy of the neutrino beam. The future measurements of this
probability at the T2K \cite{Kaneyuki:2005ze} and the subsequent experiments would shed
more lights on this model.

This paper is organized as follows:
The model and the Yukawa matrices in our model
are discussed in section 2,
and the natural realization of proton decay suppression and the 
preferable Yukawa structure are discussed in section 3.
In section 4, the predictions of our model are presented and
the impact of these solutions for T2K and lepton 
flavor violations are discussed.
The lepton asymmetry is calculated
and then it is converted into baryon asymmetry in
section 5. 
Section 6 is for the conclusion.

\section{Fermion Mass Matrices in a Minimal SO(10) Model}

We first introduce the   Yukawa interactions and the contents
of Higgs fields in the SO(10) model.
The Yukawa superpotential involves the couplings of $\bf 16$-dimensional
matter spinors $\psi_i$ ($i$ denotes a generation index) with
$\bf 10$ ($H$), $\overline{\bf 126}$ ($\overline\Delta$), and
$\bf 120$ ($D$) dimensional Higgs fields:
\begin{equation}
W_Y = \frac12 h_{ij} \psi_i \psi_j H + \frac12 f_{ij} \psi_i \psi_j
\overline\Delta
+ \frac12 h^\prime_{ij} \psi_i \psi_j D.
\end{equation}
The Yukawa couplings, $h$ and $f$ are symmetric matrices and  $h^\prime$
is an anti-symmetric matrix due to the SO(10) symmetry.
The Higgs doublet fields not only exist in $H$, $\overline\Delta$, $D$, but
 also exist in other Higgs fields which are needed in the
model. For example,
a $\bf 210$ Higgs field ($\Phi$) is employed to
 break the SO(10) symmetry down to the standard model and $\Phi$ contains  Higgs doublet fields.
One $\bf 126$ Higgs multiplet $\Delta$ is  introduced
as a vector-like pair of $\overline\Delta$ and this field also contains a Higgs doublet. The VEV of this pair
reduces the rank of SO(10) group and helps to keep supersymmetry
unbroken down to the weak scale.
Altogether, we have six pairs of Higgs doublets:
$\varphi_d = (H^{10}_d, D^{1}_d, D^{2}_d,
\overline\Delta_d, \Delta_d, \Phi_d)$,
$\varphi_u = (H_{u}^{10}, D_{u}^1, D_{u}^2,
\Delta_u, \overline\Delta_u,  \Phi_{u})$,
where superscripts $1$, $2$ of $D_{u,d}$ stand for SU(4) singlet and adjoint
pieces under the $G_{422}=\ $SU(4)$\times $SU(2)$\times
$SU(2) decomposition.
The mass term of the Higgs doublets is given as
$(\varphi_d)_a (M_D)_{ab} (\varphi_u)_b$,
and the expression of the matrix $M_D$ is given in Ref.\cite{Fukuyama:2004xs}.
The mass matrix of the Higgs doublets is diagonalized by unitary matrices
$U$ and $V$:
$U M_D V^{\rm T} = M_D^{\rm diag}$.
We assume that the lightest Higgs pair (MSSM doublets) has masses of the order of the weak scale. The MSSM Higgs doublets are given as linear combinations:
$H_d = U_{1a}^* (\varphi_d)_a$, $H_u = V_{1a}^* (\varphi_u)_a$.
Since we concentrate on the  structure of Yukawa couplings, we do not describe the dynamical reason of the mass hierarchy in
this paper.

We use ``$Y$-diagonal basis" (or SU(5) basis) to
describe the standard
model decomposition of the SO(10) representation\cite{Fukuyama:2004xs,Aulakh:2003kg}.
%
The above Yukawa interaction includes mass terms of the quark and lepton fields
as follows:
\begin{eqnarray}
W_Y^{\rm mass}
&\!\!\!=&\!\!\! h H_d^{10} (q d^c + \ell e^c) + h H_u^{10} (q u^c + \ell \nu^c) \\
&\!\!\!+&\!\!\!
          \frac1{\sqrt3}f \overline\Delta_d (q d^c -3 \ell e^c)
         + \frac1{\sqrt3}f \overline\Delta_u (q u^c - 3 \ell \nu^c)
         + \sqrt2  f \nu^c \nu^c \overline\Delta_R + \sqrt2 f \ell \ell \overline\Delta_L
 \nonumber \\
&\!\!\!+&\!\!\!     h^\prime D_d^1 (q d^c + \ell e^c) + h^\prime D_u ^1 (q u^c + \ell \nu^c) 
 +         \frac1{\sqrt3}h^\prime D_d^2 (q d^c -3 \ell e^c)
         - \frac1{\sqrt3}h^\prime D_u ^2 (q u^c -3 \ell \nu^c), \nonumber
\end{eqnarray}
where $q,u^c,d^c,\ell,e^c,\nu^c$ are the quark and lepton fields for the standard model,
which are all unified into one spinor representation of SO(10).
The VEVs of the fields $\overline\Delta_R : ({\bf 1,1},0)$ and
$\overline\Delta_L : ({\bf 1,3},1)$
give neutrino Majorana masses.
We obtain the Yukawa coupling matrices for fermions as
\begin{eqnarray}
Y_u &=& \bar h + r_2\bar f + r_3 \bar h^\prime, \label{Y_u} \\
Y_d &=& r_1(\bar h + \bar f + \bar h^\prime), \label{Y_d} \\
Y_e &=& r_1(\bar h - 3 \bar f + c_e\bar h^\prime), \label{Y_e}\\
Y_\nu &=& \bar h - 3 r_2\bar  f + c_\nu \bar h^\prime, \label{Y_nu}
\end{eqnarray}
where the subscripts $u,d,e,\nu$ denotes for up-type quark,
down-type quark, charged-lepton, and Dirac neutrino Yukawa couplings,
respectively, and
\begin{eqnarray}
\bar h &\!\!=&\!\! V_{11} h,\quad
\bar f = U_{14}/(\sqrt3\, r_1) f,\quad
\bar h^\prime = (U_{12} + U_{13}/\sqrt3)/r_1 h^\prime,\\
r_1 &\!\!=&\!\! \frac{U_{11}}{V_{11}},\quad
r_2 = r_1 \frac{V_{15}}{U_{14}}, \quad
r_3 = r_1 \frac{V_{12} -V_{13}/\sqrt3}{U_{12} + U_{13}/\sqrt3}\,, \\
c_e &\!\!=&\!\! \frac{U_{12} -\sqrt3 U_{13}}{U_{12} + U_{13}/\sqrt3},\quad
c_\nu = r_1 \frac{V_{12} +\sqrt3 V_{13}}{U_{12} +
U_{13}/\sqrt3}\,.
\end{eqnarray}
The light neutrino mass is obtained as
\begin{equation}
m_\nu^{\rm light} = M_L - M_\nu^D M_R^{-1} (M_\nu^D)^T,
\end{equation}
where $M_\nu^D = Y_\nu \langle H_u \rangle$,
$M_L = 2 \sqrt2 f \langle \overline\Delta_L \rangle$, and
$M_R = 2 \sqrt2 f \langle \overline\Delta_R \rangle$.

In this paper, we consider that the Yukawa coupling matrices for quarks and leptons
are hermitian.
The hermiticity of fermion mass matrices can be obtained in the assumption:
The original Lagrangian has symmetry under charge conjugation in addition to
the SO(10) gauge symmetry.
In the $Y$-diagonal basis, the SO(10) vector representation in terms of $X_a$
can be decomposed such that
$(X_1, X_3, X_5, X_7, X_9)$ transforms as $\bf 5$-plet,
and $(X_2, X_4, X_6, X_8, X_0)$ transforms as $\overline{\bf5}$-plet
under SU(5)$\times$U(1) decomposition.
We define the charge conjugation of the field written in the $Y$-diagonal basis,
$X_a \stackrel{C}{\leftrightarrow} X_a^*$.
The conjugation of the higher rank tensor representation is also defined similarly
using the $Y$-diagonal basis.
Since the SO(10) symmetric Lagrangian has parity invariance as an internal symmetry
(actually, it is D-parity),
the theory now has CP symmetry.
Imposing that Lagrangian is invariant under the CP conjugation,
the Yukawa couplings, $h_{ij}$, $f_{ij}$ and $h^\prime_{ij}$ and
all masses and couplings in the Higgs superpotential are all real.
However, when SO(10) symmetry is broken down,
the CP symmetry can be spontaneously broken by the VEV of $\bf 45$ Higgs field.
Due to the non-existence of cubic terms for $\bf 45$ Higgs field,
the VEV of $\bf 45$ can be pure imaginary.
Consequently,
the mixing of the lightest Higgs doublets with the Higgs doublets present
in {\bf 120} involves a pure imaginary coefficient
which will make the fermion masses hermitian
 in this model\cite{Dutta:2004hp}.
This symmetry has wider implications. For instance, the
model presents a solution to the SUSY CP problem and strong CP problem.
It is possible to explain the quark masses and mixing angles and the neutrino sectors by using the
above parameters\cite{Dutta:2004hp}.

\section{Proton Decay and Flavor Structure}
As mentioned above, it was shown recently by us\cite{Dutta:2004zh} 
that the suppression of proton decay for all
tan$\beta$ determines the flavor structure of the three matrices
$h,f,h'$ to a very narrow range. We review this argument in this
section.

The proton decay is mediated by the colored Higgs triplets,
$\varphi_T+ \varphi_{\overline T}$ : ($({\bf 3},{\bf 1},-1/3)+ c.c.$) and
 $\varphi_C + \varphi_{\overline C}$ : ($({\bf 3}, {\bf 1}, -4/3)+ c.c.$).
These Higgs triplets appear in $\bf 10$+$\bf 120$+$\bf 126$+$\overline{\bf 126}$+$\bf 210$ multiplets.
We generate both $LLLL$ ($C_L$) and $RRRR$ ($C_R$)
operators:
\begin{equation}
-W_5 =\frac12 C_L^{ijkl} q_k q_l q_i \ell_j + C_R^{ijkl} e_k^c u_l^c
u_i^c d_j^c .
\end{equation}
These operators are obtained by integrating out the triplet Higgs fields,
$\varphi_{\overline T} = (H_{\overline T},D_{\overline T},D_{\overline
T}^\prime,\overline
\Delta_{\overline T},\\ \Delta_{\overline T},\Delta_{\overline T}^\prime,
\Phi_{\overline T})$
and
$\varphi_T$ =$\ (H_{T},D_{T},D_{T}^\prime,\Delta_{T},\overline\Delta_{T},
\overline\Delta_{T}^\prime,  \Phi_{T})$. 
The fields with `$^\prime$'
are decuplet, and the others are sextet or {\bf 15}-plet under SU(4) decomposition.
The $C_R$ operator 
 has also contributions from the triplets, 
$\varphi_{\overline C} = (D_{\overline C}, \Delta_{\overline C})$
and $\varphi_C = (D_C, \overline\Delta_C)$.
The mass term of the Higgs triplets are
given as
$(\varphi_{\overline T})_a (M_T)_{ab} (\varphi_T)_b +
(\varphi_{\overline C})_a (M_C)_{ab} (\varphi_C)_b$.
The mass matrices, $M_T$ and $M_C$, are 7$\times$7 and 2$\times$2 matrices
respectively, and their explicit forms are given in the
literature\cite{Fukuyama:2004xs}.
The Yukawa couplings which cause proton decay are written as
\begin{eqnarray}
W_Y^{\rm trip.} &\!\!\!=&\!\!\! h H_{\overline T}\ (q \ell + u^c d^c)
+ h H_{T}\ (\frac12 qq + e^c u^c )
 +f \overline\Delta_{\overline T}\, (q \ell -  u^c d^c)
+   f\overline\Delta_{T}\, (\frac12 qq -  e^c u^c ) \nonumber\\
&\!\!\!+&\!\!\!  {\sqrt2} f \overline\Delta_{T}^\prime\ e^c u^c
+{\sqrt2} h^\prime (D_{\overline T}\ u^c d^c
+ D_{\overline T}^\prime\ q \ell -
  D_{T}\ e^c u^c
+  D_{T}^\prime\ e^cu^c)  \nonumber \\
&\!\!\!+&\!\!\!  2 f \overline\Delta_{C} \, d^c e^c + 2 h^\prime
D_{\overline C} \ u^c u^c
+2 h^\prime D_{C} \ d^c e^c.
\end{eqnarray}

The dimension five operators are
written by the Yukawa couplings $h$, $f$ and $h^\prime$ as follow:
\begin{eqnarray}
C_L^{ijkl} &=& c \, h_{ij}h_{kl}+ x_1 f_{ij}f_{kl} +
x_2 h_{ij}f_{kl} + x_3 f_{ij}h_{kl}
 +
x_4 h^\prime_{ij}h_{kl} + x_5 h^\prime_{ij}f_{kl} , \label{LLLL} \\
 C_R^{ijkl}
&=& c \, h_{ij}h_{kl}+ y_1 f_{ij} f_{kl} + y_2 h_{ij}f_{kl} + y_3
f_{ij}h_{kl} + y_4 h^\prime_{ij}h_{kl}  + y_5 h^\prime_{ij}f_{kl}
 \nonumber
\\
&+& y_6  h_{ij}h^\prime_{kl} + y_7  f_{ij}h^\prime_{kl} + y_8
h^\prime_{ij} h^\prime_{kl} + y_9 h^\prime_{il} f_{jk} + y_{10} h^\prime_{il} h^\prime_{jk}. \label{RRRR}
\end{eqnarray}
The coefficient $c$ is given as $c=(M_T^{-1})_{11}$,
and the other coefficients $x_i$, $y_i$ are also given by the components of
$M_T^{-1}$ or $M_C^{-1}$.
Note that the $H_T$ and $\overline\Delta_T$ have
opposite D-parity and we get $y_3 = - x_3$.
The $y_9$ and $y_{10}$ terms are generated by $\varphi_C + \varphi_{\overline C}$.

The proton decay operators can be written conveniently
by diagonalizing the Higgs triplet mass matrix $M_T$ by two unitary matrices,
$X$ and $Y$,
as $X M_T Y^{\rm T} = {\rm diag} (M_1, M_2, \cdots, M_7)$,
\begin{eqnarray}
 C_L^{ijkl}
&\!\!\!=&\!\!\! \sum_a \frac1{M_a}(X_{a1} h + X_{a4} f + \sqrt2 X_{a3} h^\prime)_{ij}
(Y_{a1} h + Y_{a5} f)_{kl}
,\label{LLLL3}\\
 C_R^{ijkl}
&\!\!\!=&\!\!\! \sum_a \frac1{M_a} (X_{a1} h - X_{a4} f + \sqrt2 X_{a2} h^\prime)_{ij}
(Y_{a1} h - (Y_{a5}- \sqrt2 Y_{a6}) f 
+ \sqrt2
(Y_{a3}-Y_{a2}) h^\prime )_{kl} \nonumber \\
&&+(y_9, y_{10}\ {\rm terms}) \label{RRRR3}.
\end{eqnarray}
In the SU(5) limit, only one of the colored triplets is much lighter
than the others, i.e. $M_1 \ll M_a$ $(a\neq 1)$,
and we can obtain the
following relations for the diagonalizing matrices from the
explicit form of the Higgs mass matrices in Ref.\cite{Fukuyama:2004xs,Aulakh:2003kg}:
$U_{11}=X_{11}$, $V_{11}=Y_{11}$, $U_{14}=X_{14}=0$,
$V_{15}: Y_{15} : Y_{16} = \sqrt3 : 1 : \sqrt2\,$,
$U_{12} : U_{13} : X_{12} : X_{13} = V_{12} : V_{13} : Y_{12} : Y_{13}
= 1 : \sqrt3 : \sqrt2 : \sqrt2\,$.
Using the above relations, we get
\begin{equation}
r_2 \rightarrow \infty, \ \bar f \rightarrow 0, \ r_3 = 0, \ c_e = -1, \
c_\nu=2 r_1 V_{12}/U_{12}, \label{su5}
\end{equation}
for the Yukawa matrices in Eqs.(\ref{Y_u}-\ref{Y_nu})
and thus, as expected, we get the SU(5) relations,
$Y_u = Y_u^{\rm T}$,
$Y_d = Y_e^{\rm T}$,
and the dimension five proton decay operators can be written in terms of
the Yukawa couplings as
$C_L^{ijkl} \simeq C_R^{ijkl} \simeq (Y_d)_{ij} (Y_u)_{kl}/M_1$ and we know that proton decay is not suppressed at this
limit and this limit is also not good to satisfy the quark-lepton masses.

The proton decay amplitude can be written
as $A = \alpha_2 \beta_p/(4\pi M_T m_{SUSY}) \tilde A$, where
\begin{equation}
\tilde A = c\tilde A_{hh}+ x_1\tilde A_{ff}+x_2\tilde A_{hf}+
x_3\tilde A_{fh} +x_4\tilde A_{h^\prime h}+ x_5\tilde A_{h^\prime f} .
 \label{tildea}\end{equation}
 The coefficients $c$ and $x_i$ are given in Eq.(\ref{LLLL}),
and there are also similar $C_R$ contributions.
To satisfy the current nucleon decay bounds, we need
$|\tilde A_{p\rightarrow K\bar\nu}| \alt 10^{-8}$,
$|\tilde A_{n\rightarrow \pi\bar\nu}| \alt 2 \cdot 10^{-8}$
and $|\tilde A_{n\rightarrow K\bar\nu}| \alt 5 \cdot 10^{-8}$
if the colored Higgsino mass is $2 \cdot 10^{16}$ GeV, and squark and
wino masses are
around 1 TeV and 250 GeV, respectively. Instead of inducing any cancellation among
different terms in the amplitude, we will try to suppress the individual contributions.

One way to suppress the decay amplitude is by demanding
cancellation among different terms, a strategy common in the
literature.
In order to achieve that, we need a cancellation among $h$, $f$ and
$h^\prime$ to have small couplings for first and second
generations in the expressions in Eqs.(\ref{LLLL},\ref{RRRR}).
 However, we also need cancellation among the same couplings to generate the large mass hierarchy among the
quark masses and in general, the coefficients $r_2, \,r_3$ in up-type Yukawa matrix
and $x_i$, $y_i$ in proton decay operators are
unrelated. Further,
the  $\overline{\bf 126}$ Higgs contribution
has  opposite signatures ($y_3 = - x_3$)
for  $C_L$ and $C_R$. Therefore, the
cancellation required to obtain small Yukawa coupling for $Y_u$ by
tuning $r_2 \bar f$ 
can not simultaneously suppress both
$C_L$ and $C_R$ operators by tuning $X_{14}$ in Eqs.(\ref{LLLL3}, \ref{RRRR3}).
Moreover,  the  $\bf 120$ contribution to the
 proton decay amplitude has vanishing contribution to
the $kl$ part of $C_L^{ijkl}$ due to antisymmetric $h'$.
Thus, if the cancellation in $Y_u$ requires a tuning of $r_3
h^\prime$,  the
$C_L$ operator can not be suppressed.

The best way to avoid the cancellation is to choose smaller values of $r_{2,3}$.
Let us start with $r_{2,3}\simeq 0$. In this case, the $C_L$ can be written as
$C_L^{ijkl} \propto (Y_u + \gamma h^\prime)_{ij}(Y_u)_{kl}$
and in the operator $C_R^{ijkl}$,
$kl$ part is also related to $Y_u$.
This will correspond to the case where $X_{14},Y_{15}
\sim 0$.
The $RRRR$ contribution to $p \rightarrow K \bar\nu_\tau$ mode
is  suppressed  compared to the minimal SU(5) model by a suppression factor $\lambda_u/\lambda_d\sim 1/100$ for
$\tan\beta\sim 50$.
Similarly, since the $kl$ part of $C_L$ are also related to the
$Y_u$ instead of $Y_d$, the $LLLL$ contribution to the $p
\rightarrow K \bar\nu$ is also suppressed even for
$\tan\beta \sim 50$, compared to the SU(5) model (since
$\lambda_c/\lambda_s\sim1/5$). However, these
suppressions are not enough to satisfy the current experimental bound.

In order to satisfy the bounds naturally, we need $\tilde A_{hh} \alt 5
\cdot 10^{-8}$ in the expression, Eq.(\ref{tildea}).
If $\tilde A_{hh} \agt 10^{-7}$,
we need to tune $x_i$ and $y_i$ for every decay mode to cancel $\tilde
A_{hh}$, which is unnatural.
The $\tilde A_{hh}$ depends on the magnitudes of the elements
from the [1,2] block of $\bar h$ which is
specifically determined from the fit to the up-type Yukawa coupling
as a function of $r_2$ and $r_3$.
In the case $r_{2,3} \sim O(1)$, the suppression of
up- and charm-quark Yukawa couplings are acquired by fine-tuning of $r_{2,3}$,
and thus the elements from [1,2] block of $\bar h$ are of the order of
down- and strange-quark Yukawa couplings.
In that case, we find $\tilde A_{hh}$  $ \sim 10^{-4}$ which requires
a very high level of fine-tuning for all the decay modes.
In the case $r_2, r_3 \sim 0$ for generical fits,
the Yukawa coupling $\bar h$
is close to $Y_u$ in the $Y_u$-diagonal basis by definition.
However, even in the case,
the $\tilde A_{hh}$ is of the order of $10^{-7}$ for generical fits
and further tuning among the
coefficients $x_i$ and $y_i$ is needed to satisfy the current experimental data.
We therefore need a specific type of
Yukawa texture to suppress the proton decay rate.
%
To suppress $\tilde A_{hh}$, the elements $\bar h_{11}$ and $\bar h_{22}$
(in $\bar h$-diagonal basis)
are needed to be suppressed rather than the up- and charm-quark Yukawa couplings,
respectively.
As a result, we need Yukawa texture to be
$\bar h \simeq $ diag$(\sim 0,\sim 0,O(1))$.
Once $\bar h$ is fixed, the other matrices $\bar f$ and $\bar h^\prime$
are almost determined as
\begin{equation}
\bar f \simeq \left(\begin{array}{ccc}
\sim 0 & \sim 0 & \lambda^3 \\
\sim 0 & \lambda^2 & \lambda^2 \\
\lambda^3 & \lambda^2 & \lambda^2
\end{array}
\right) , \quad
\bar h^\prime \simeq i \left(\begin{array}{ccc}
0 & \lambda^3 & \lambda^3 \\
-\lambda^3 & 0 & \lambda^2 \\
-\lambda^3 & -\lambda^2 & 0\end{array}
\right),
\label{special_texture}
\end{equation}
where $\lambda \sim 0.2$.
The correct charm mass is generated 
by $r_2 m_s/m_b \simeq \lambda_c$ ($|r_2| \simeq 0.1-0.15$), and
down-quark mass and Cabibbo angle $\theta_C$ are generated
by $\bar h^\prime_{12}$
with $m_d/m_s \simeq \sin^2\theta_C$.
Then we need $\bar f_{11} \alt O(\lambda^6)$, $\bar f_{12} \alt O(\lambda^4)$
and $r_3 \alt O(\lambda^2)$ to
obtain proper size of up-quark and electron masses.
We have $r_1\simeq m_b/m_t\tan\beta$.
 In the basis where $Y_u$ is diagonal, $\tilde
A_{hh}$ in this texture is not completely zero but can become much smaller than
$10^{-8}$.

We show one example for numerical fit for $\tan\beta (M_Z)= 50$ :
$\bar h = {\rm diag} (0,0, 0.638)$,
\begin{equation}
\bar f \simeq \left(\begin{array}{ccc}
0 & -0.00044 & 0.00208 \\
-0.00044 & 0.00945 & 0.0101 \\
0.00208 & 0.0101 & 0.0071
\end{array}
\right) , \quad
\bar h^\prime \simeq i \left(\begin{array}{ccc}
0 & -0.0022 & 0.00046 \\
0.0022 & 0 & 0.0181 \\
-0.00046 & -0.0181 & 0\end{array}
\right),
\end{equation}
$r_1= 0.966$, $r_2 = 0.135$, $r_3=0$, $|c_e| = 0.987$.

After we  suppress the  $\tilde A_{hh}$,
we also need to examine the contribution of the
 other components e.g. $\tilde A_{ff,hf,fh,h'f,h'h,\cdots}$.
Their coefficients, $x_i, y_i$, involve the
colored Higgs mixings, which can be suppressed by our choice of
the vacuum expectation values and the Higgs couplings.
According to our numerical studies, some of the mixing angles must be less than
about a few percent in the case of $\tan\beta \sim 50$ to suppress the decay.
However, the mixing angles can become larger as $\tan\beta$ becomes smaller.
In the above example of numerical fit,
$p \rightarrow K \bar\nu_\mu$ and $p \rightarrow K \bar\nu_\tau$
modes are dominant for $LLLL$ operator,
and for $RRRR$ operator, respectively.
The $\tilde A_{hh}$ for $p \rightarrow K \bar\nu_\mu$ mode is $\sim 2\cdot 10^{-11}$.
The amplitudes for other components are $10^{-8}-10^{-6}$.

We have seen that the suppression of proton decay requires $r_3\simeq 0$,
which is same as SU(5) condition for $\bf 120$ Higgs coupling,
and $r_2\simeq 0.1 - 0.15$, which however is not a SU(5) condition.
%
The second condition, as well as suppression of colored Higgs mixing, is implemented by
requiring $U_{14} \gg V_{15},\,X_{14},\,\,Y_{15}$.
%
We note that if $U \simeq V$ is satisfied (which is like a up-down symmetry
and therefore $\tan\beta \simeq 50$),
the $r_3 \simeq 0$ condition also satisfies
the SU(5)-like condition 
\begin{equation}
c_e \simeq -1\,, \qquad c_\nu \simeq 2,
\label{ce-cnu}
\end{equation}
for {\bf 120} Higgs coupling in $Y_e$ and $Y_\nu$,
though these
conditions are not required to be satisfied to fit the fermion masses or to suppress the
proton decay.

\section{The Model Predictions}

Now we are ready to discuss the predictions of the model. The
number of parameters in the models is 17: 3($h$), 6($f$), 3($h'$)
and 5 Higgs parameters ($r_{1,2,3}$, $c_{e,\nu}$). (We are
working in the diagonal $h$ basis). Now in the process of
explaining the proton decay, some of the parameters are redundant
to fit masses and mixings. We choose $\bar h_{11,22}=0$ and
$r_3=0$. The $r_3$ can be $O(\lambda^2)$ in our Yukawa texture,
but such a value of $r_3$ gives only a small correction to the
CKM mixings.
 We use a very small
$f_{11}$, but even in the case $\bar f_{11}=0$, the obtained
Yukawa texture can fit all masses and mixings. We make it a free
parameter since it affects the leptonic fitting. Since we will be
working type II seesaw (actually, the Yukawa texture of $\bar f$
is compatible to bi-maximal mixings), $c_\nu$ is redundant in
fitting fermion masses and mixings. This reduces the number of
parameters to 13.
In
order to fix the remaining 13 parameters, we use the up-type quark
masses, charged lepton masses, the CKM angles and the phase, the
ratio of the  squared of neutrino mass differences ($\Delta
m^2_{\rm sol}/\Delta m^2_{A}$), and the bi-maximal mixings as
input parameters. Consequently, the down-type quark masses,
$U_{e3}$ and $\delta_{\rm MNSP}$ etc are the predictions of this
model.

We itemize below the essential features of our predictions.

\begin{figure}[t]
\center
 \includegraphics[width=9cm]{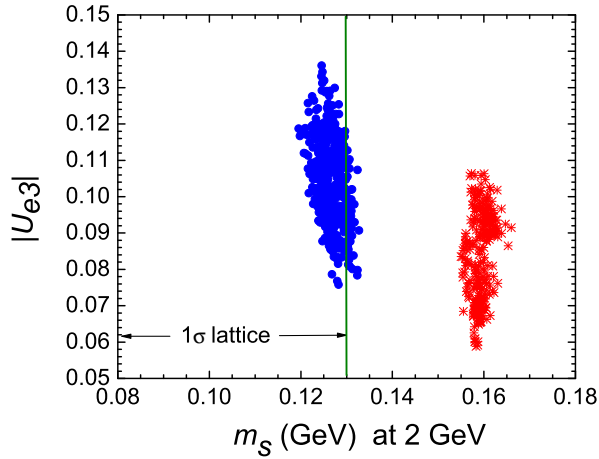}%
 \caption{\label{msue3} $|U_{e3}|$ is plotted as a function of $m_s$.
 The two regions are described in the text.}
 \end{figure}
\begin{figure}[t]
 \center
 \includegraphics[width=8cm]{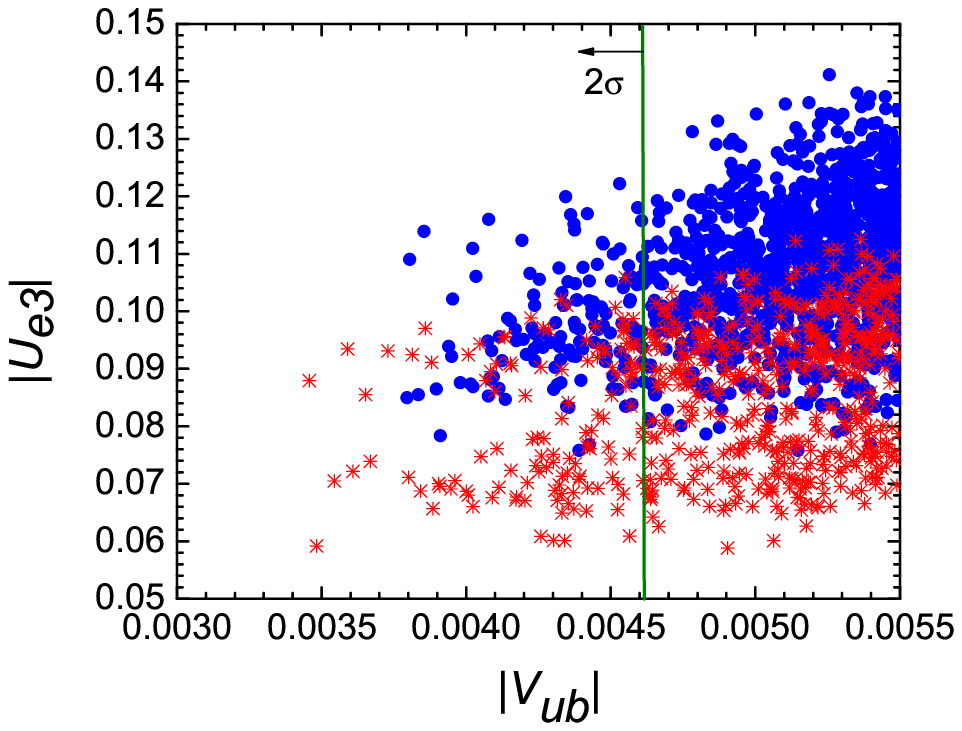}%
%
%
 \includegraphics[width=8cm]{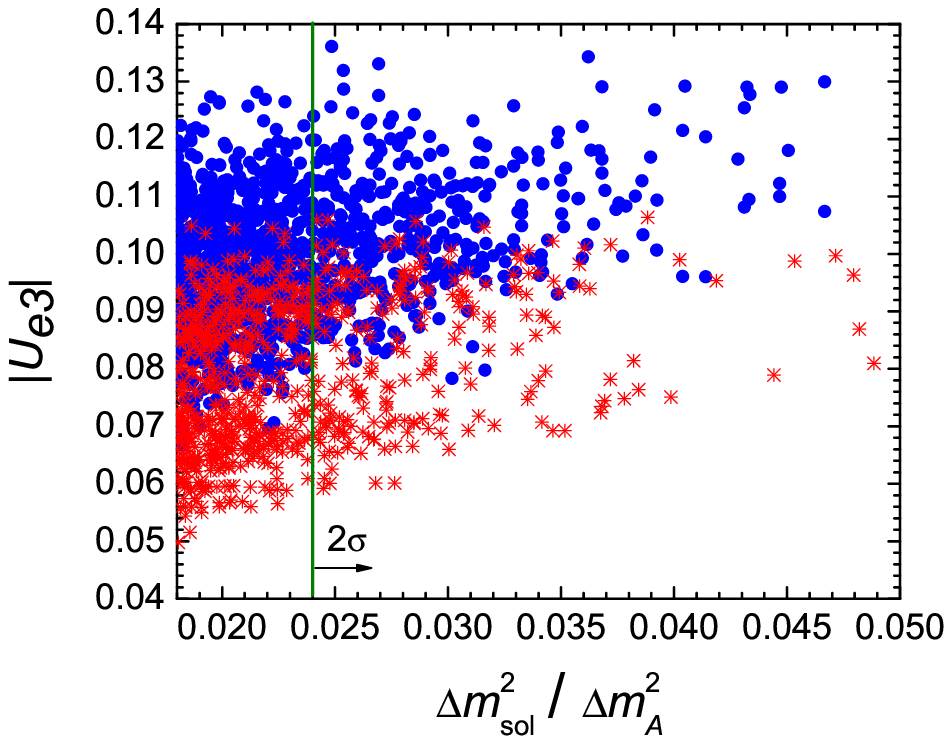}%
 \caption{\label{sqratioue3} $|U_{e3}|$ is plotted as a function of
$V_{ub}$ and
$\Delta m^2_{\rm sol}/\Delta m^2_{A}$. }
\end{figure}

\subsection{Strange Quark Mass}
In order to see how the down quark masses and in particular the
strange quark mass gets constrained in our model, we first note
from Eqs.(\ref{Y_d},\ref{Y_e},\ref{special_texture}) we get
$\det M_e \simeq |c_e|^2 \det M_d\,$.,
when $\bar f_{11} \ll \lambda^5$ and thus,
 the down quark and
the charged lepton mass matrix leads to
 the Georgi-Jarskog (GJ) relation naturally when the
SU(5) relation ($c_e = -1$) is satisfied in $\bf 120$ Higgs
mixing.

The strange quark mass is predicted since no cancellation happens
to derive $V_{cb}$ (due to $r_2, r_3 \ll 1$ and the hermiticity
condition). The predicted value of strange quark mass has two
separate regions, roughly $m_s \sim 1/3\, m_\mu (1 \pm
O(\lambda^2))$. In our  numerical calculations, the negative sign
corresponds to a  strange quark mass: $\overline m_s (\mu=2\,
{\rm GeV}) \sim 120-130$ MeV. (The strange quark mass at 1 GeV is
obtained by multiplying the strange mass at 2 GeV with 1.35.)
This strange mass value is in the range of lattice derived value,
$ \overline m_s (\mu=2\, {\rm GeV}) = (105 \pm 25)\ {\rm
MeV}$\cite{Eidelman:2004wy}. The GJ relation is realized for this
result and thus the fitting of the lattice result supports $|c_e|
\simeq 1$ (when $\bar f_{11} \ll \lambda^5$). If we use the
positive signature in the rough estimation,
 we find the following value of the strange
quark mass, $\overline m_s (\mu=2\, {\rm GeV}) \sim 155-165$ MeV
which is allowed by the QCD calculation\cite{Eidelman:2004wy} (In
this case, non-zero values of $\bar f_{11}$ are needed to make
$|c_e|=1$). 
In Figure \ref{msue3}, we plot  $|U_{e3}|$ against the
strange quark mass. We see the two regions corresponding to two
different values of strange quark masses (corresponding to two
different signs). We also note that the larger values of strange
mass prefers lower values of $U_{e3}$. We use $\alpha_s (M_Z) =
0.118$ and a 3-loop QCD and 1-loop QED running to evolve the
strange mass from 2 GeV scale to the electroweak
scale\cite{Tarasov:1980au}.

Our prediction of strange-down quark mass ratio is $m_s/m_d \simeq 17-18 $, $19-20.5$.
The non-lattice value of the ratio is $m_s/m_d = 18.9 \pm 0.8$\cite{Leutwyler:2000cw}.

The $m_u/m_d$ ratio is not a prediction of this model since $m_u$ is an input.
To obtain a large atmospheric mixing, $\bar f_{33}$ needs to be suppressed,
and thus bottom-tau Yukawa coupling needs to be unified within several percent.
In order to achieve such a situation, large $\tan\beta \sim 50$ is needed
(or $\tan\beta \sim 2$). Thus, the bottom mass is predicted.  However, this prediction is lost since it gets a sizable contribution from the soft SUSY breaking terms
for large $\tan\beta$.




\subsection{$|U_{e3}|$}
Since there is no cancellation, we get the following stable approximate relation for $U_{e3}$:
\begin{eqnarray}
|U_{e3}|^2 &\approx& \frac{\tan^2\theta_{\rm sol}}{1-\tan^4 \theta_{\rm sol}}
\Delta m^2_{\rm sol}/\Delta m^2_{A}\,. \end{eqnarray}
We also have
the following relation since $U_{e3}$ is related to the ratio $\bar f_{13}$ and $\bar f_{23}$:
\begin{eqnarray}
|U_{e3}| &\approx& \frac1{\sqrt2} \left|\frac{V_{ub}}{V_{cb}}\right| .
\end{eqnarray}
These relations are obtained since $\bar f_{12}$ is small compared to $\bar f_{13}$.
Actually, when $\bar f_{12}\simeq \bar f_{13}$, $U_{e3}$ can be canceled to be zero
 without relating to the solar mixing angle and $V_{ub}$.
In our proton decay suppressed texture,
the $\bar f_{12}$ is suppressed due to the assumption $\bar h_{11} \ll \lambda_u$,
and thus the $|U_{e3}|$ prediction can be stable.

\begin{figure}[t]
 \center
 \includegraphics[width=9cm]{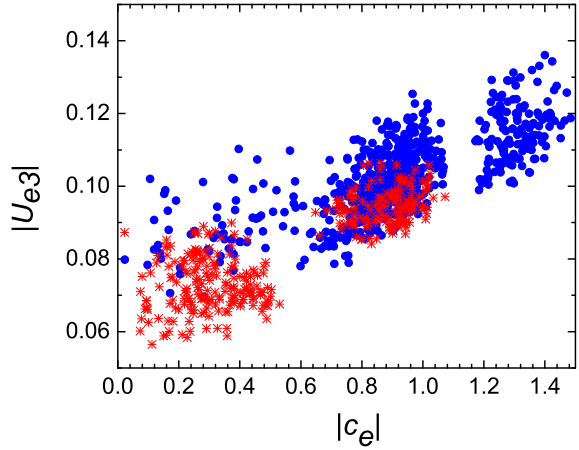}%
 \caption{\label{ue3ce} $|U_{e3}|$ is plotted as a function of $c_e$. }
\end{figure}

The $|U_{e3}|$ is bounded from below through the mass squared
ratio and solar mixing angle and has an upper limit due to the
experimental bound on $|V_{ub}| = (3.67\pm0.47)\times 10^{-3}$.
In Figure~\ref{sqratioue3}, we plot $|U_{e3}|$ as a function
$V_{ub}$ and $\Delta m^2_{\rm sol}/\Delta m^2_{A}$. We find that
for $|V_{ub}|<0.0045$ and $\Delta m^2_{\rm sol}/\Delta m^2_{A} >
0.022$, the $|U_{e3}|$ is bounded to be $0.06 - 0.11$.
We  plot $|U_{e3}|$ as a function $|c_e|$ in
Figure~\ref{ue3ce}. The value of $c_e=-1$ (corresponding to the SU(5) condition for $\bf 120$)
gives rise to $|U_{e3}|\sim 0.1$. Both large and small strange quark masses are allowed for 
$c_e=-1$.

\begin{figure}[t]
 \includegraphics[width=8cm]{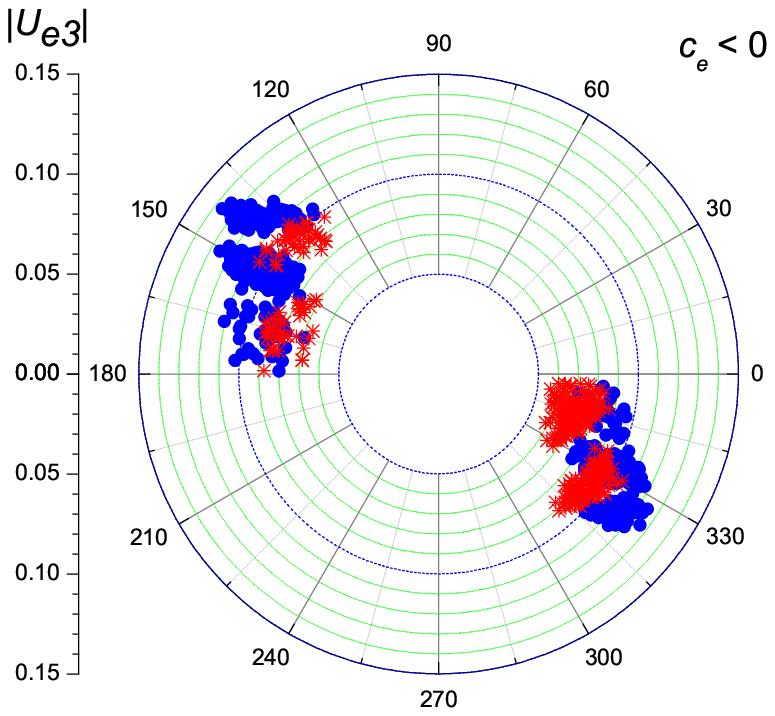}%
%
 \includegraphics[width=8cm]{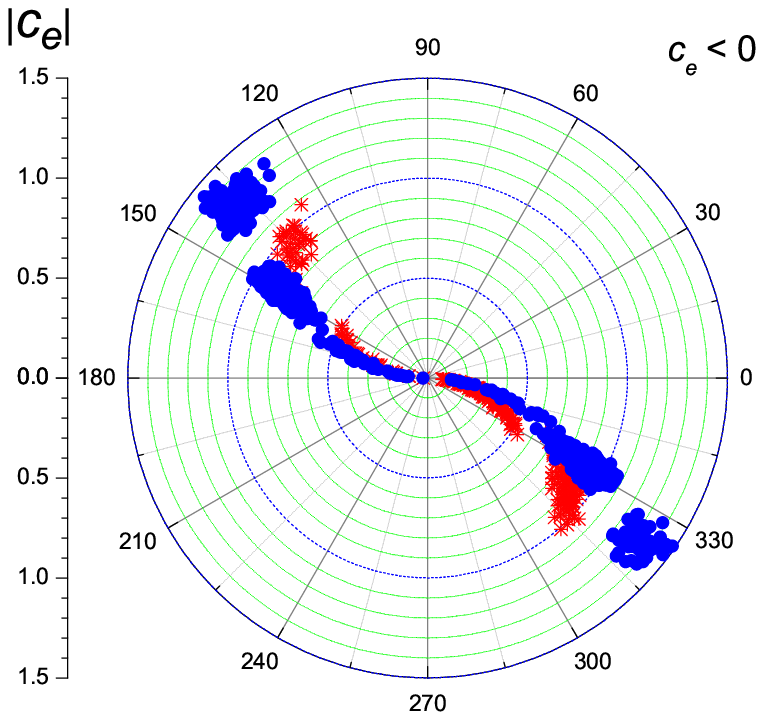}%
 \caption{\label{mnspue3} $|U_{e3}|$ and $c_e$ is shown as a function of $\delta_{\rm MNSP}$ for $c_e<0$. }
\end{figure}

\subsection{MNSP Phase}

The MNSP phase is also stable in this model and is given by the approximate expression:
\begin{equation}
\sin \delta_{\rm MNSP} \sim \frac1{\sqrt2} \frac{\sin \theta_{12}^e}{\sin \theta_{13}^\nu}
\sin \left(\tan^{-1} \frac{c_e \bar h^\prime_{12}}{3 \bar f_{12}}\right),
\end{equation}
where $\theta_{12}^e$ and $\theta_{13}^\nu$
are mixing angles in the diagonalizing matrix of $Y_e$ and neutrino mass matrix, respectively.
Approximately, $\sin \theta_{13}^\nu \simeq |U_{e3}|$. We plot $|U_{e3}|$  as a function of
MNSP phase $\delta_{\rm MNSP}$ in
Figure~\ref{mnspue3}. We find that the $\delta_{\rm MNSP}$ lies in the 2nd or 4th quadrant
if $c_e < 0$. The 1st and the 3rd quadrants are absent since  the solar mixing angle
becomes small due to a  cancellation between
 $\theta_{12}^\nu$ and $\theta_{12}^e$.
In Figure~\ref{mnspue3}, we plot $|c_e|$ as a function of $\delta_{\rm MNSP}$.
We see that $c_e = -1$ (SU(5) relation of $\bf 120$ Higgs mixing,
and also with a favorable value by the GJ relation), generates
$\sin \delta_{\rm MNSP} \simeq \pm (0.5-0.7)$.
To obtain this result, we assume that type II contribution is dominant.



\begin{figure}[t]
 \center
 \includegraphics[width=10cm]{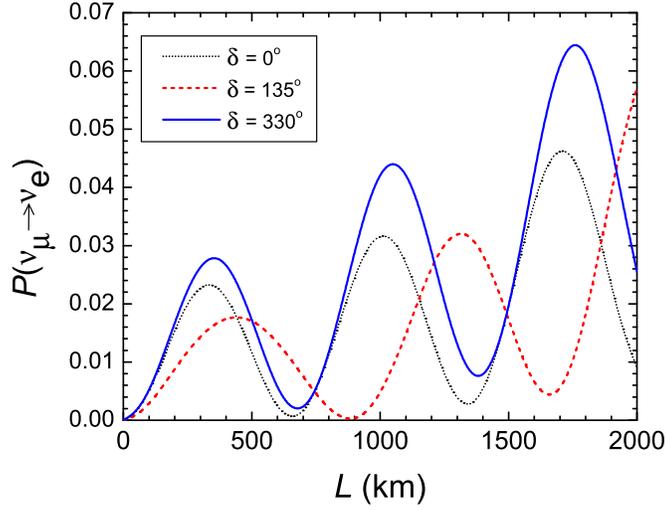}%
 \caption{\label{numunue}The probability $P_{\nu_\mu \rightarrow \nu_e}$ as a function of $L$(km).}
\end{figure}

 The location  of $\delta_{\rm MNSP}$ in the 2nd or 4th quadrant
has impact on the probability of $\nu_{\mu}$ to $\nu_e$ oscillation
 ($P_{\nu_\mu \rightarrow \nu_e}$) which will be measured at the T2K experiment and
at the newly proposed Tokai-to-Korea
 experiment\cite{Hagiwara:2005pe}.   This probability depends on sine and
cosine of  $\delta_{\rm MNSP}$, distance ($L$), energy of the neutrino beam, mass squared
differences ($\Delta m^2_{13}$, $\Delta m^2_{12}$), 3 mixing angles, and matter density.
We take the energy of the beam is 0.7 GeV and
the values of  
mass squared differences:
$\Delta m^2_{13}=2.5\times10^{-3}\, {\rm eV}^2$,
$\Delta m^2_{12}=8\times10^{-5}\, {\rm eV}^2$ and $U_{e3}=0.1$. In Figure~\ref{numunue}, we plot
the probability as a function of distance.
The probability for $\delta=330^{\rm o}$ is about 1.8 times bigger compared to
the probability for $\delta=135^{\rm o}$
 when the beam arrives at Kamioka from Tokai ($L=295$ km).
The difference is magnified much more if we have  a detector  installed
at Korea ($L=1000$ km). Also we notice that a peak in the distribution appears at T2K and at the
 Tokai-to-Korea experiment. The location of the peak however will change if we change the above parameters.
%

\section{Leptogenesis}

Let us consider the calculation of baryon asymmetry of the Universe. 
As we will see below, a lepton asymmetry is generated by CP violating decays 
of right-handed neutrinos out of equilibrium\cite{Fukugita:1986hr}. 
The sphaleron processes\cite{Kuzmin:1985mm} convert the lepton asymmetry 
to a baryon asymmetry\cite{Buchmuller:2004nz}. 
Our goal is to see if the recent
experimental value of the baryon-to-photon ratio, i.e.
\cite{Tegmark:2003ud}
\begin{equation}
\eta_B = (6.3 \pm {0.3}) \times 10^{-10},
\label{WMAP}
\end{equation}
can be reproduced in our model. 


The baryon-to-photon ratio is obtained as
\begin{equation}
\eta_B = - a \frac1{f_*^0} \kappa \,\epsilon_1 \,,
\label{wmap}
\end{equation}
where $a$ is a sphaleron conversion factor $a\simeq 0.35$, and
$1/f_*^0$ is coming from the fact that
the photon number density increases as the degrees of freedom
present at the epoch of leptogenesis annihilate and is given by
$f_*^0=g^*/g^0$, where $g^*$ is the number of degrees of
freedom at $T=M_{N_1}$ (mass of the lightest right-handed neutrino)
and $g^0=3.91$, giving $1/f_*^0\simeq 0.016$.
The $\epsilon_1$ is the amount of CP violating lepton asymmetry via 
the lightest right-handed neutrino,
which is defined as
\begin{equation}
\epsilon_1 = \sum_i
\frac{\Gamma(N_1 \rightarrow \ell_i H^*_u)
 -\Gamma(N_1 \rightarrow \bar\ell_i H_u)}{
\Gamma(N_1 \rightarrow \ell_i H^*_u)
 +\Gamma(N_1 \rightarrow \bar\ell_i H_u)}\,.
\end{equation}
The $\kappa$ is an efficiency factor, which does not related on the
CP violation of lepton asymmetry and parameterizes the effects of
scattering and decay processes.
In the thermal leptogenesis scenario,
the $\kappa$ is a calculable number by solving Boltzman equation.
The $\kappa$ is a function of an effective neutrino mass, $\tilde m_1$,
which defined as
\begin{equation}
\tilde m_1 = \frac{[(\hat M_\nu^D)^\dagger \hat M_\nu^D]_{11}}{M_{N_1}},
\end{equation}
where $\hat M_\nu^D$ is the Dirac neutrino mass matrix in the basis 
where right-handed Majorana mass is diagonal with real and positive eigenvalues, $M_{N_i}$.
The Ref.\cite{Nielsen:2001fy} shows that
the $\kappa$ for $10^{-2}$ eV $< \tilde m_1 < 10^3$ eV is approximately given
as 
\begin{equation}
\kappa \simeq 0.3 \left(\frac{10^{-3} {\rm eV}}{\tilde m_1}\right)
\left(\ln \frac{\tilde m_1}{10^{-3} {\rm eV}}\right)^{-0.6} \,,
\end{equation}
and in Ref.\cite{Buchmuller:2004nz}, the $\kappa$ is given by a power law as
\begin{equation}
\kappa = (2\pm1) \times 10^{-2} \left(\frac{0.01\ {\rm eV}}{\tilde m_1}\right)^{1.1\pm 0.1}.
\end{equation}

Now let us see the prediction of our model.
As we have seen, flavor structure has been already fixed in our model,
and thus the $\epsilon_1$ and $\kappa$ can be calculated as a function of
the lightest right-handed neutrino mass and SU(2)$_L$ triplet Higgs mass.
For example, 
since the $[(\hat M_\nu^D)^\dagger \hat M_\nu^D]_{11}$ is almost fixed
in our model, the effective neutrino mass is a function of the lightest
right-handed neutrino, $M_{N_1}$.
As a result, the efficiency factor $\kappa$ depends on only $M_{N_1}$. 
Consequently, the recent experimental data, Eq.(\ref{WMAP}), gives us
a prediction to the mass scale.

Let us discuss the possible contributions to the CP violating lepton asymmetry 
in our model.
The decay amplitude of the lightest right-handed neutrino in one-loop
involves the right-handed neutrinos and  SU(2)$_L$ triplet Higgs\cite{Hambye:2003ka} 
present in ${\bf 126}\,+\,\overline{\bf 126}$ and $\bf 54$ 
(if there is one). 
The right-handed neutrino loop contribution is obtained as
\begin{equation}
\epsilon_1^N = -\frac1{8\pi} \sum_{i=2,3} \frac{{\rm Im}\, 
[(\hat Y_\nu^\dagger \hat Y_\nu)_{1i}{}^2]}
{(\hat Y_\nu^\dagger \hat Y_\nu)_{11}} F\left(\frac{M_{N_i}^2}{M_{N_1}^2}\right),
\label{epsilon_1}
\end{equation}
where $\hat Y_\nu$ is the Dirac neutrino Yukawa coupling in the basis where
right-handed Majorana mass is diagonal with real and positive eigenvalues, $M_{N_i}$.
Since the expression of the $\epsilon_1^N$ does not depend on the overall scale of
right-handed Majorana neutrino mass,
the value of $\epsilon_1^N$ is almost determined in our model.
The value is proportional to $c_\nu$, which is a coefficient of $\bf 120$
Higgs contribution to Dirac neutrino Yukawa coupling, Eq.(\ref{Y_nu}).
Our prediction is \cite{fuku}
\begin{equation}
|\epsilon_1^N| = (1-4)\times 10^{-4} \,(c_\nu/2)\,.
\label{type-I-epsilon}
\end{equation}
Suppose that this $\epsilon_1^N$ dominate the lepton asymmetry $\epsilon_1$,
the value of $\kappa$ is determined to be $\kappa \simeq (0.25-1) \times 10^{-3}\,(2/c_\nu)$
to satisfy the current experimental data.
The values of $\kappa$ correspond to
the effective neutrino mass, $\tilde m_1 = (0.1 - 0.4)\, (c_\nu/2) $ eV,
and we obtain the lightest right-handed Majorana mass
is
$M_{N_1} = (0.4-1) \times 10^{13}\, (2/c_\nu)$ GeV.
Using the relations, $f = \sqrt3 r_1/U_{14} \bar f$, $M_R = 2\sqrt2 f v_R$,
we obtain the corresponding value of VEV is
$v_R \simeq U_{14}\,(2/c_\nu)\, (1-2.5) \times 10^{15}$ GeV.

We also have the triplet Higgs loop contribution, $\epsilon_1^\Delta$,
and the contribution may disturb the above prediction of lightest right-handed neutrino mass.
In fact, 
as discussed in Ref.\cite{Hambye:2003ka,Antusch:2004xy},
this triplet Higgs loop contribution may dominate $\epsilon_1$ naively
when we consider type II seesaw.
However, the amount of lepton asymmetry depends on the flavor structure,
and such naive estimation may not be correct.
Actually, in our model, the $\epsilon_1^\Delta$ can be calculated
as a function of lightest right-handed Majorana mass,
and the triplet mass.
To see this, let us see how we can realize type II seesaw.

In order to satisfy the type II VEV magnitude for $v_L 
\equiv \langle \overline\Delta_L \rangle$,
 one $\bf 54$ Higgs is needed\cite{Goh:2004fy}. 
There are then two pairs of SU(5) submultiplet
 {\bf 15}s in the theory ({\bf 15} Higgs of SU(5) is the one that contains the
 triplet that leads to the type II seesaw).
  For the triplet VEV term to dominate,
one of these two pairs must have a mass around a scale of
$10^{13}$ GeV or so. It was shown in Ref.\cite{Goh:2004fy} that
 a linear combination of the SU(5) {\bf 15}
 sub-multiplets in {\bf 54} and {\bf 126} can have a mass around $10^{13}$
 GeV without conflicting with the symmetry breaking and coupling unification and the other
 one becomes closer to the SO(10) breaking scale.

 We can write down the  superpotential involving the triplets and their interactions
 (for simplicity, we omit the terms from $\bf 210$ Higgs):
\begin{equation}
W = \lambda_1 H_u H_u \Delta_E + \lambda_2 v_R \Delta_E \overline \Delta_L
+ \bar\lambda_2 v_R \overline\Delta_E \Delta_L
+ Y_\Delta LL \overline \Delta_L + M_\Delta \Delta_L \overline \Delta_L
+ M_E \Delta_E \overline \Delta_E \,,
\end{equation}
where $Y_\Delta = 2\sqrt2 f$,
and $\Delta_E$ is a triplet in {\bf 54} Higgs.
The VEV is given as
\begin{equation}
v_L \simeq \lambda_1 s_\Delta c_\Delta \frac{v_u^2}{M_{\Delta_1}} \,,
\end{equation}
where $s_\Delta$, $c_\Delta$ are mixing (sin and cos) of SU(2)$_L$ triplets,
\begin{equation}
s_\Delta c_\Delta \simeq \frac{\lambda_2 v_R}{M_{\Delta_2}} \,.
\end{equation}
We denote $M_{\Delta_{1,2}}$ as eigenmasses of SU(2)$_L$ triplets
and we assume 
$M_{\Delta_1}\ll M_{\Delta_2}$ to make type II dominant
and $M_\Delta$ to be around GUT scale.
Note that $M_{\Delta_1}$ is a free parameter even if we fix the $v_L$.
However, assuming that the Higgs couplings are less than $O(1)$,
the $M_{\Delta_1}$ should be less than around $10^{13}$ GeV.

The lepton asymmetry via triplets in the loop  is given in the 
Ref.\cite{Hambye:2003ka,Antusch:2004xy},
\begin{eqnarray}
\epsilon_1^\Delta &=&
- \frac{3}{8\pi} \lambda_1 c_\Delta s_\Delta \frac{M_{N_1}}{M_\Delta}
\frac{{\rm Im} (Y_\nu^\dagger Y_\Delta Y_\nu^*)_{11}}{(Y_\nu^\dagger Y_\nu)_{11}}
\,G(y)\\\nonumber
&=&- \frac{3}{8\pi} \frac{M_{N_1}}{v_u^2}
\frac{{\rm Im} (Y_\nu^\dagger m_\nu^{\rm II} Y_\nu^*)_{11}}{(Y_\nu^\dagger Y_\nu)_{11}}
\,G(y)\,,
\end{eqnarray}
where $G(y)= y \ln \frac{y+1}{y}$ and $y = M_{\Delta_1}^2/M_{N_1}^2$.
Our prediction of the triplet loop contribution is
\begin{equation}
|\epsilon_1^\Delta| = (2.5 - 6) \times 10^{-4} \ (c_\nu/2)
\left( \frac{M_{N_1}}{10^{13}\ {\rm GeV}} \right)
\left( \frac{m_{\nu_3}^{\rm II}}{0.05\ {\rm eV}} \right) G(y)\,.
\label{type-II-epsilon}
\end{equation}
We note that the function $G(y)$ is always smaller than 1.
If the triplet mass $M_{\Delta_1}$ is 10\% compared to the lightest right-handed 
Majorana mass, $M_{N_1}$, we obtain $G(y)\simeq 0.046$
and the triplet loop contribution $\epsilon_1^\Delta$ can be negligible compared to
the right-handed neutrino loop contribution $\epsilon_1^N$.
When $M_{\Delta_1}$ and $M_{N_1}$ are comparable,
then the both triplet and right-handed neutrino loops contribute
to the lepton asymmetry.

It is possible to check the predictions, Eqs.(\ref{type-I-epsilon},\ref{type-II-epsilon}), 
using the Yukawa couplings shown as an example in section 3.
the Dirac neutrino coupling in the above basis with $c_\nu =2$ 
(which is a favorable value as we noted in the last paragraph in section 3) is given by:
\begin{equation}
\hat Y_{\nu} = \left(\begin{array}{ccc}
0.002 & 0.003\,\exp(-1.54\,i) & 0.0026\,\exp(-0.344\,i) \\
-0.0167 & 0.021\,\exp(-1.53\,i) & 0.025\,\exp(3.37\,i)  \\
-0.229& 0.417\,\exp(4.70\,i)& 0.422\,\exp(-3.019\,i)
\end{array}
\right).
\end{equation}
The Majorana neutrino coupling matrix which gives the right-handed
neutrino masses is proportional to the $f$ coupling,
\begin{equation}
{\hat{\overline f}} = {\rm diag.} (0.00108,0.00303,0.0185)\,.
\end{equation}

If there is no huge cancellation between $\epsilon_1^N$ and $\epsilon_1^\Delta$,
the resulting lepton asymmetry $\epsilon_1$ needs to be $O(10^{-4})$ 
and consequently, the efficiency factor $\kappa$ is $O(10^{-3})$
and the lightest right-handed Majorana mass, $M_{N_1}$, is $10^{13}$ GeV.
Note that we also have solutions where the Majorana mass scale is larger.
In this case, since $\kappa$ becomes larger,
the lepton asymmetry $\epsilon_1$ needs to be canceled
between $\epsilon_1^N$ and $\epsilon_1^\Delta$
by choosing the triplet mass.


We note that the triplet Higgs decay also can produce lepton asymmetry by its
own decay, but this contribution
 vanishes when the Hermitian structure of Yukawa coupling is assumed.

\begin{figure}[t]
\center
 \includegraphics[width=8cm]{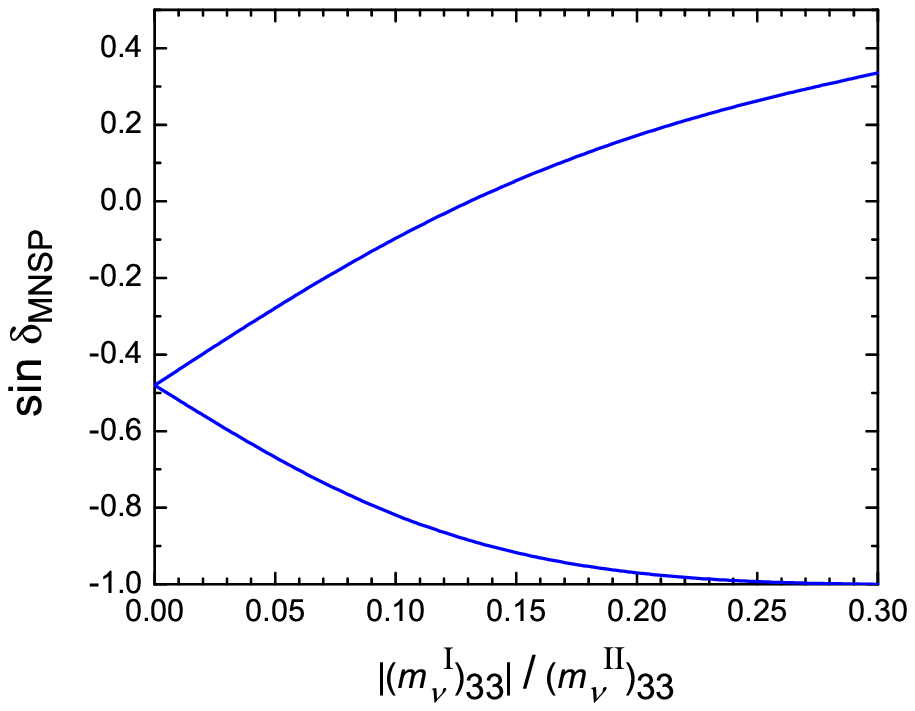}%
 \includegraphics[width=8cm]{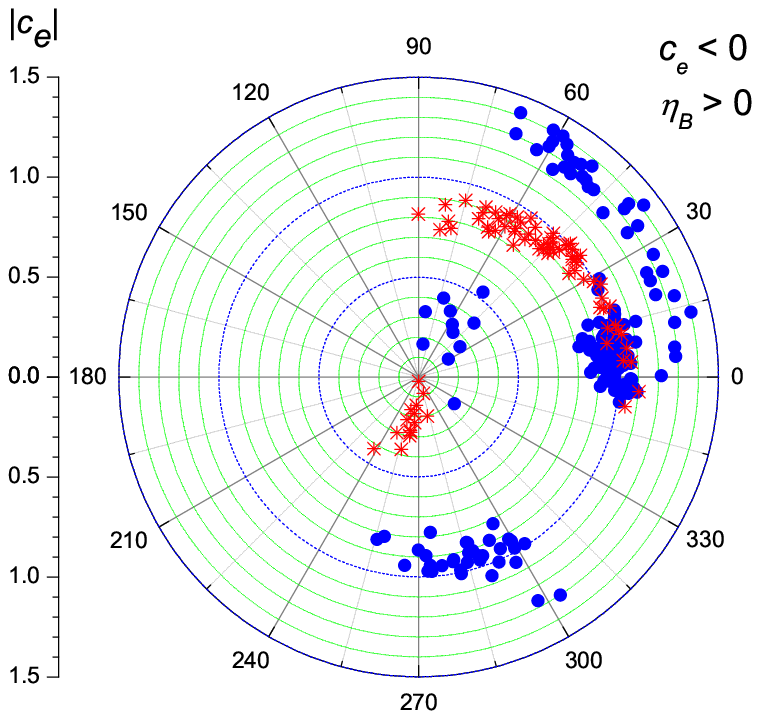}%
 \caption{\label{sindelta-split}$\sin \delta_{\rm MNSP}$ is plotted as a function of the
 ratio of the type I contribution to the type II
 contribution ($|(m_{\nu}^{\rm I})_{33}|/(m_{\nu}^{\rm I\!I})_{33}$) 
 in the heaviest neutrino mass[Left].
The  $|c_e|$ is plotted as a function of $\delta_{\rm
 MNSP}$ for $\eta_B>0$ and [Right] for $v_R \simeq U_{14}(2/c_\nu)\ (1-2.5) \times 10^{15}  $GeV. }
 \end{figure}


When $M_{N_1} \sim 10^{13}$ GeV,
the type I contribution can disturb the prediction of oscillation parameters
obtained in the previous section.
We find that the $|U_{e3}|$ prediction is not changed much,
but the prediction of the MNSP phase is modified due to the phase of type I contribution.
As shown in figure \ref{sindelta-split},
the $\sin \delta_{\rm MNSP}$ splits from the type II prediction
when type I contribution increases (However, the correct neutrino fit in our model
can not be obtained by pure type I contribution). In figure \ref{sindelta-split}, we show the
$\sin \delta_{\rm MNSP}$  as a function of the
 ratio of the type I contribution to the type II
 contribution ($|(m_{\nu}^{\rm I})_{33}|/(m_{\nu}^{\rm I\!I})_{33}$) 
 to the heaviest  neutrino mass.
We use the ${\bar h}$, ${\bar f}$ and
 ${\bar h'}$ values presented in section 3 (we use $c_e\simeq-1$ and $c_\nu=2$) and
vary $v_R/U_{14}$ continuously.
In figure \ref{sindelta-split},
we also show the MNSP phase for the case of positive $\eta_B$ dominated by the contribution
given in Eq.(\ref{epsilon_1}).
When $v_R / U_{14} \agt 2 \times 10^{16}$ GeV,
type I contribution can be negligible
and the MNSP phase lies around 135$^{\rm o}-150^{\rm o}$, 315$^{\rm o}-330^{\rm o}$
(when $c_e \simeq -1$)
as we have seen in the previous section.
In this scenario of large $v_R$, the generated lepton asymmetry 
 may be an order of magnitude bigger compared to the observation since 
 the efficiency factor $\kappa$ becomes larger.
However, the left-handed triplet $\Delta_L$ contribution in the loop can produce a
10\% level cancellation and generate the right amount of lepton
asymmetry.
The location of the MNSP phase, which can be observed at T2K and the subsequent
experiments as mentioned before,
can be a probe to distinguish the scenario.


In our discussion so far, we did not concern ourselves with the
nature of SUSY breaking mechanism. As is well known, 
gravitino production puts an upper
bound of about $10^9$ GeV on the reheating temperature $T_R$ after
inflation when the gravitino mass is in the range, 100 GeV $\alt m_{3/2} \alt$ 1 TeV\cite{kim}. 
Since in our model, the lightest right-handed
neutrino which is supposed to be responsible for leptogenesis has
a mass of $10^{13}$ GeV or so, clearly the thermal leptogenesis
picture outlined in this section will not work in this
case. 
The possibilities to make the thermal leptogenesis scenario available
are to consider
a light gravitino such as $m_{3/2} < 16$ eV\cite{Viel:2005qj}, 
late time entropy production in a gauge mediation model\cite{Fujii:2002fv},
quasi thermal picture\cite{Allahverdi:2005fq}.
Otherwise, one can consider a non-thermal leptogenesis where the inflaton plays a key
role\cite{shafi}.
%
%
We do not give details of the implications of such a
choice for our SO(10) model except to note that it does not
affect the fermion sector of the model.

\section{Lepton Flavor Violation}

We now discuss the lepton flavor violating processes e.g. $\mu\rightarrow
e\gamma$, $\tau\rightarrow
\mu\gamma$ etc.
The operator for $l_i\rightarrow l_j+\gamma$ is:
\begin{equation}\label{eq501}
{\cal L}_{l_i\rightarrow l_j\gamma} = \frac{i e}{2 m_l} \, \overline{l_j}
\,\sigma^{\mu\nu} q_{\nu} \left( a_l P_L + a_r P_R \right) l_i
\cdot A_\mu + h.c.
\end{equation}
where $P_{L,R} \equiv (1 \mp \gamma_5)/2$ and $\sigma^{\mu\nu}
\equiv \textstyle{\frac{i}{2}} \,[ \gamma^{\mu}, \gamma^{\nu} ]$.
The decay width for $l_i\rightarrow l_j+\gamma$ can be written as:
\begin{equation}\label{eq502}
\Gamma(l_i\rightarrow l_j+\gamma) = \frac{m_\mu e^2}{64\pi} \left( |a_l|^2 +
|a_r|^2 \right).
\end{equation}
Then the branching ratio is obtained by multiplying this decay width with the
lifetime of the $l_i$ lepton.
The supersymmetric contributions include the neutralino and
chargino diagrams.

\begin{figure}[t]
 \center
 \includegraphics[width=7.5cm]{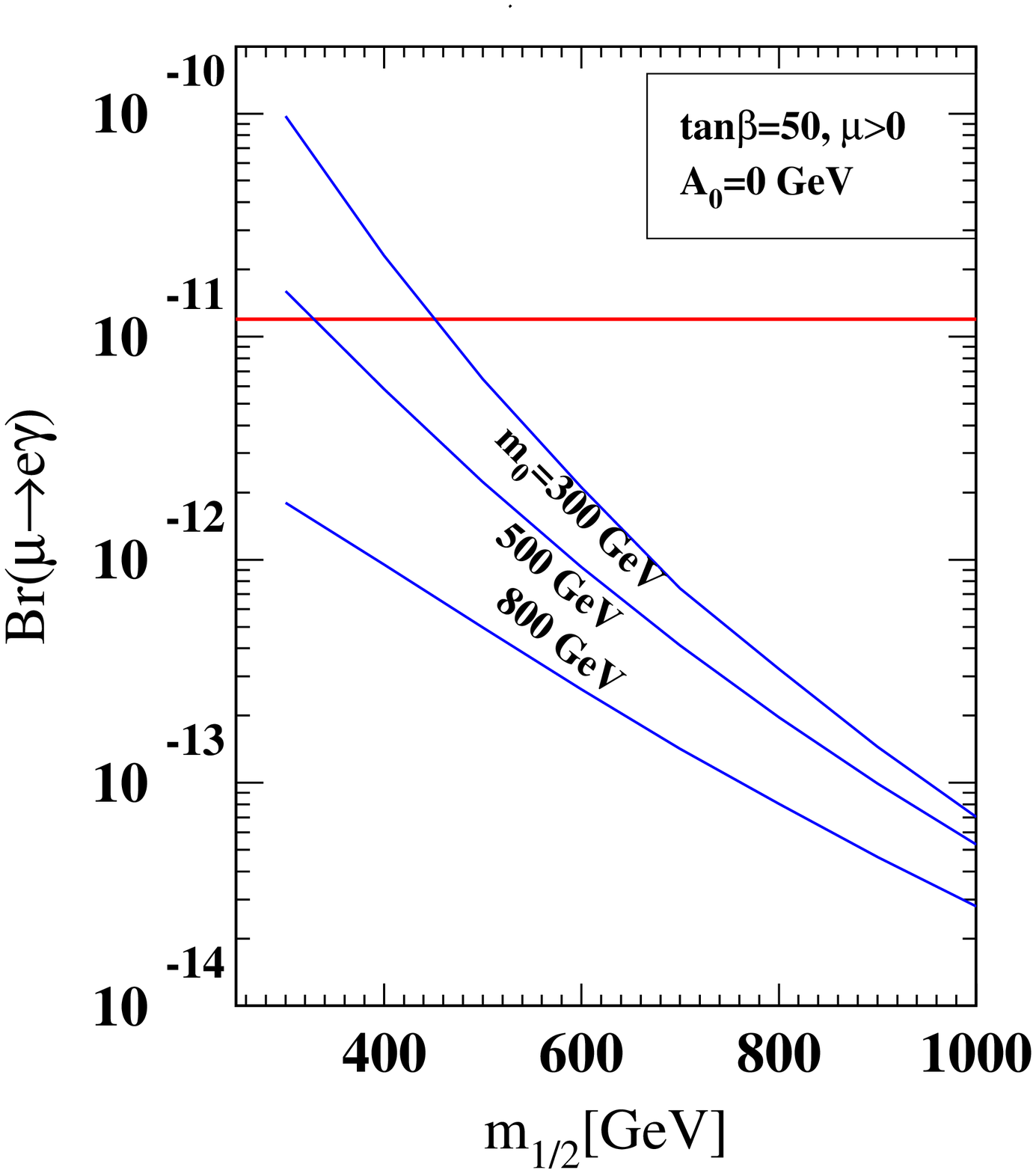}%
 \includegraphics[width=7.5cm]{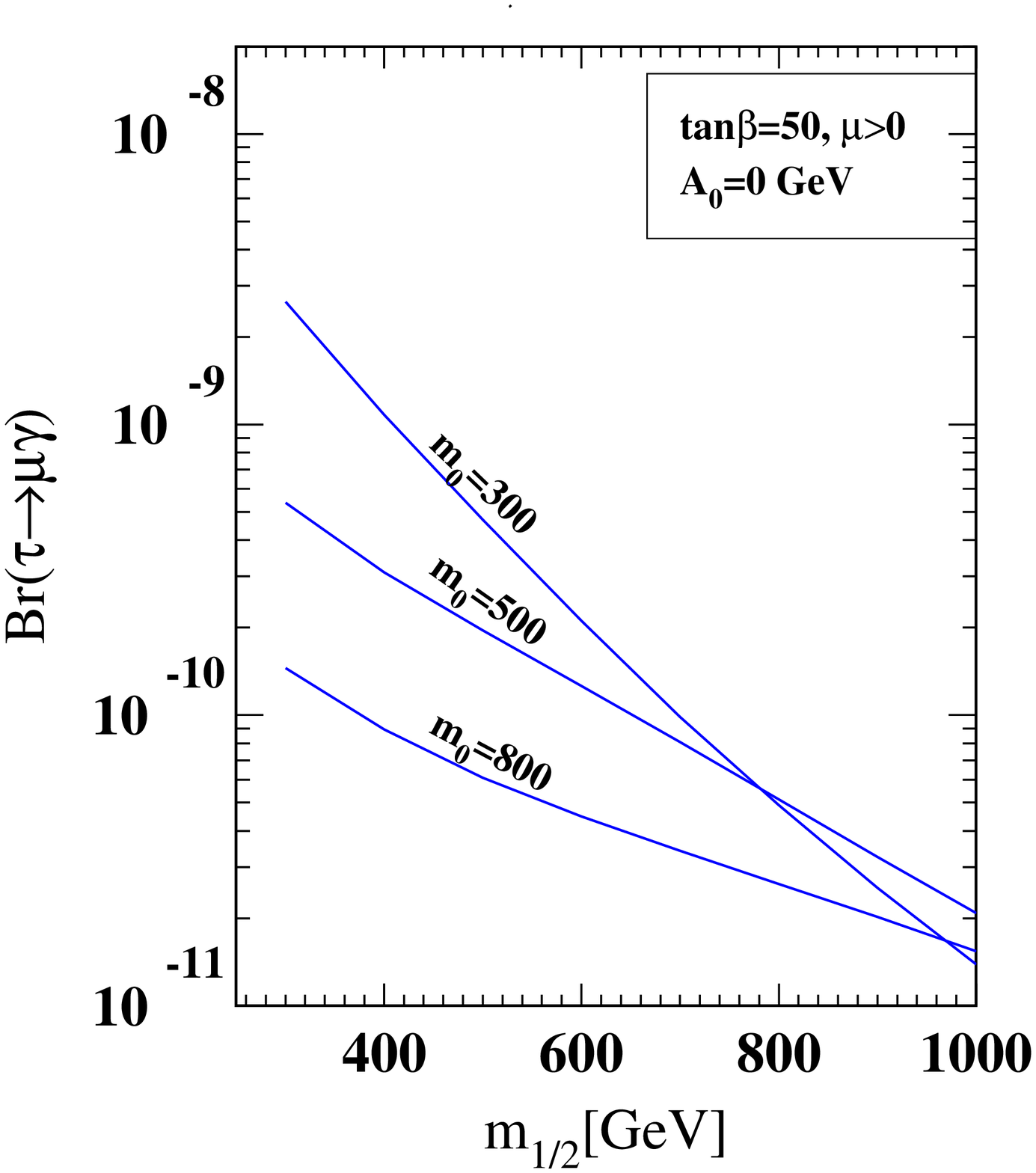}%
 \caption{\label{muegm} BR[$\mu\rightarrow e+\gamma$] and BR[$\tau\rightarrow \mu+\gamma$]
   are plotted as a function of $m_{1/2}$ for different values of $m_0$. }
 \end{figure}

We work in the basis where the charged lepton masses are
diagonal at the highest scale of the theory. We use the mSUGRA  boundary conditions,
i.e. $m_0\,$: universal scalar mass, $m_{1/2}\,$: universal gaugino mass, $A_0\,$: universal
trilinear mass term,  $\tan\beta$, sign of $\mu$ to calculate the BRs. Since we have
$\lambda_b\simeq\lambda_{\tau}$, the masses of the Higgs are not unified with the
universal scalar mass for the sfermions for realistic parameter space after including
the finite SUSY one loop correction to the b-quark mass.
We plot BR[$\mu\rightarrow e+\gamma$] and BR[$\tau\rightarrow \mu+\gamma$]
as a function of $m_{1/2}$ for different values of
 $m_0$ in Figure~\ref{muegm}, 
 and we find that the BR[$\mu\rightarrow e+\gamma$] can be quite large.

\section{Conclusion}

The suppression of proton decay is a major challenge for
the grand unified models. We have constructed a minimal SO(10) model which can suppress the
proton decay naturally. The model has $\bf 10$, $\overline{\bf 126}$ and $\bf 120$ Higgs
multiplets to generate the fermion masses. The CP symmetry of the model keeps the fermion
masses hermitian at the grand unified scale and the SUSY CP problem is also under control. The
neutrino masses are generated by type II seesaw.
The model has 13 parameters to fit quark and lepton masses and mixings,
and thus it gives predictions for down-type quark
masses, $|U_{e3}|$ and $\delta_{\rm MNSP}$. 
The prediction for $|U_{e3}|$ is $0.06-0.11$ 
with the upper
limit being imposed by $V_{ub}$ and the lower limit by the mass squared ratio
($\Delta m_{\rm sol}^2/\Delta
m_{A}^2$) and the solar
mixing angle. The strange quark mass can be  small (lattice calculated size) or
large (QCD calculated size). The phase $\delta_{\rm MNSP}$ is  in the 2nd or
 4th quadrant when the SU(5)-like condition is satisfied in the 120 coupling.
This feature has distinctive impact on the  probability of $\nu_{\mu}$ to $\nu_e$ oscillation
 ($P_{\nu_\mu \rightarrow \nu_e}$) which will be measured at the T2K and subsequent experiments.
The lepton asymmetry generated in the decay of the right-handed neutrinos in
 this model produces the observed amount of baryon asymmetry.
Depending on the magnitude of the symmetry breaking scale $v_R$,
 the observed baryon asymmetry predict the $\delta_{\rm MNSP}$ phase in different quadrants.
The lepton flavor violating Br[$\mu\rightarrow e+\gamma$] can be large in this model.

\section*{Acknowledgments}
This work of B.D. and Y.M. is supported by
the Natural Sciences and Engineering Research Council of Canada and
the work of R. N. M. is supported by the National Science Foundation
Grant No. PHY-0354401. We also thank A. Mazumdar for helpful comments.

\end{document}